\documentclass[journal]{IEEEtran}
\usepackage{multirow}
\usepackage{amssymb}
\usepackage{amsmath}
\usepackage{amsfonts}
\usepackage{dsfont} 
\usepackage[utf8]{inputenc}
\usepackage{amsthm}
\usepackage{cite}
\usepackage{subfigure}
\usepackage{color}
\usepackage{times}
\usepackage{graphicx}
\usepackage{psfrag}
\usepackage[dvipsnames]{xcolor}
\usepackage{enumitem}

\makeatletter
\newcommand*\rel@kern[1]{\kern#1\dimexpr\macc@kerna}
\newcommand*\widebar[1]{%
  \begingroup
  \def\mathaccent##1##2{%
    \rel@kern{0.8}%
    \overline{\rel@kern{-0.8}\macc@nucleus\rel@kern{0.2}}%
    \rel@kern{-0.2}%
  }%
  \macc@depth\@ne
  \let\math@bgroup\@empty \let\math@egroup\macc@set@skewchar
  \mathsurround\z@ \frozen@everymath{\mathgroup\macc@group\relax}%
  \macc@set@skewchar\relax
  \let\mathaccentV\macc@nested@a
  \macc@nested@a\relax111{#1}%
  \endgroup
}
\makeatother

\newcommand{\bm}{\mathbf}
\newcommand{\be}{\begin{equation}}
\newcommand{\ee}{\end{equation}}
\newcommand{\bea}{\begin{eqnarray}}
\newcommand{\eea}{\end{eqnarray}}

\newcommand{\bA}{{\bm A}}
\newcommand{\bI}{{\bm I}}

\newcommand{\bD}{{\bf D}}

\newcommand{\bP}{{\bf P}}

\newcommand{\bH}{{\bf H}}

\newcommand{\BDFT}{{\boldsymbol{\mathcal F}}}

\IEEEoverridecommandlockouts
\begin{document}

\title{Downlink Transmission in FBMC-based Massive MIMO with Co-located and Distributed Antennas}

\author{\normalsize Hamed Hosseiny$^\dagger$, Arman Farhang$^*$, and Behrouz Farhang-Boroujeny$^\dagger$  
\\$^\dagger$ECE Department, University of Utah, USA, \\
$^*$Department of Electronic and Electrical Engineering, Trinity College Dublin, Ireland.  \\
Email: \{hamed.hosseiny, farhang\}@utah.edu, \{arman.farhang\}@tcd.ie}
\maketitle

\begin{abstract}
\color{black}
This paper introduces a practical precoding method for the downlink of Filter Bank Multicarrier-based (FBMC-based) massive multiple-input multiple-output (MIMO) systems. The proposed method comprises a two-stage precoder, consisting of a fractionally spaced prefilter (FSP) per subcarrier to equalize the channel across each subcarrier band. This is followed by a conventional precoder that concentrates the signals of different users at their spatial locations, ensuring each user receives only the intended information. In practical scenarios, a perfect channel reciprocity may not hold due to radio chain mismatches in the uplink and downlink. Moreover, the channel state information (CSI) may not be perfectly known at the base station. To address these issues,  we theoretically analyze the performance of the proposed precoder in presence of imperfect CSI and channel reciprocity calibration errors. Our investigation covers both co-located (cell-based) and cell-free massive MIMO cases. In the cell-free massive MIMO setup, we propose an access point selection method based on the received SINRs of different users in the uplink. Finally, we conduct numerical evaluations to assess the performance of the proposed precoder. Our results demonstrate the excellent performance of the proposed precoder when compared with the orthogonal frequency division multiplexing (OFDM) method as a benchmark.
\color{black}

\end{abstract}

\vspace{-4mm}
\section{Introduction}
The success of massive multiple-input multiple-output (MIMO) technology in the recent roll-out of the fifth generation wireless systems (5G) is an advocate on the importance of multiple antenna techniques for future networks \cite{giordani2020toward}. Thus, massive MIMO will be among the key building blocks that underpin the future of 5G Advanced and the sixth generation wireless networks (6G) \cite{saad2019vision}. The shortcomings of orthogonal frequency division multiplexing (OFDM) such as its high sensitivity to synchronization errors were taken on board in the design of 5G new radio (5G NR) standard by the introduction of the flexible subcarrier spacings \cite{5gsystem}. However, OFDM still suffers from bandwidth efficiency loss considering the extended cyclic prefix of the length $25\%$ of symbol duration, \cite{5gsystem}. Furthermore, the advent of new applications, such as autonomous driving, where wireless channels become highly time varying, call for alternative waveforms that are more resilient than OFDM to the time variations of the channel \cite{schulz2017latency}. Filter bank multicarrier (FBMC) is one of a kind with a high bandwidth efficiency and resilience to the synchronization errors and the channel time variations \cite{farhang2011ofdm,nissel2017filter, aminjavaheri2015impact}.

\color{black}
 The above observations, clearly, justify the significance of exploring FBMC-based massive MIMO as a candidate technology for the future wireless systems. FBMC-based massive MIMO was first introduced in \cite{farhang2014}. In this work, the authors showed how FBMC benefits from the channel flattening effect of massive MIMO to widen the subcarrier bands and thus further improve bandwidth efficiency. As a follow up contribution, in \cite{farhang2014pilot}, the authors addressed the pilot contamination problem in the uplink of FBMC-based Massive MIMO systems. Further studies in \cite{rottenberg2018performance} and \cite{singh2019uplink} provide the mean squared error (MSE) and sum-rate performance of FBMC in the uplink of massive MIMO channels, respectively. Channel estimation and equalization aspects of FBMC-based massive MIMO were covered in \cite{hosseiny2021fbmc,9508836} and \cite{aminjavaheri2015frequency,aminjavaheri2018filter}, respectively. While ideal scenarios are considered in a large body of the available literature on the topic, in a more recent work, we focused on the practical aspects of FBMC-based massive MIMO systems \cite{hosseiny2021fbmc}. In particular, we investigated imperfect channel state information (CSI) effects in both co-located (cell-based), and distributed  (cell-free) antenna setups \cite{hosseiny2021fbmc}. 

\textcolor{black}{Cell-free massive MIMO was introduced to address some shortcomings of co-locating antennas in massive MIMO, such as the low service quality of the users at the vicinity of cell edges \cite{ngo2017cell}. The downlink of cell-free massive MIMO was studied for the narrowband systems in the literature \cite{interdonato2019downlink,ngo2017total,buzzi2018user}. Authors in \cite{interdonato2019downlink} considered the application of downlink training for a cell-free massive MIMO. They showed that due to the limited channel hardening, it is beneficial to use downlink training. The antenna selection that yields the user-centric scenario of cell-free massive MIMO was studied in \cite{buzzi2018user}. Two antenna selection methods, one based on the received signal power and the other based on the large-scale fading coefficient, were proposed. In \cite{ngo2017total}, the spectral efficiency of the cell-free massive MIMO and its limitations were derived.} \color{black}
Many emerging applications in future wireless networks require ultra-reliable low-latency communications (URLLC), \cite{schulz2017latency}. 
The features of FBMC make it a promising candidate for multi-user distributed systems and asynchronous communication in a delay-stringent setup, \cite{aminjavaheri2015impact}. The resilience to synchronization errors makes FBMC a good fit for distributed multi-user systems where perfect synchronization is challenging, \cite{aminjavaheri2015impact}. Thanks to lower overhead and the ability to widen subcarrier bands in FBMC,  FBMC-based distributed antenna systems are able to satisfy low-delay distributed antenna requirements.

\color{black}
Among  variants of FBMC, such as FBMC with offset QAM (FBMC-OQAM), FBMC-QAM, and complex-based FBMC-OQAM (C-FBMC-OQAM), FBMC-OQAM is examined in this work. FBMC-OQAM, which is also known as staggered multi-tone (SMT) \cite{farhang2011ofdm}, separates real and imaginary parts of each QAM symbol and transmits them with a time offset of half symbol interval. To allow transmission of QAM symbols, FBMC-QAM makes use of a pair of prototype filters. To avoid interference among different symbols, both across time and frequency, the second prototype filter leads to a design with a very poor out-of-band leakage; e.g., see filter responses in Fig. 12 of \cite{nam2016new}. This leads to a significant loss of the benefits that FBMC can offer, \cite{chung2014synchronization}, \cite{khrouf2018much}.
In C-FBMC-OQAM, the same basis functions as FBMC/OQAM are used, however, QAM symbols with half of the power are transmitted in two FBMC  symbols \cite{kong2020frame}. Assuming approximately flat gain over each subcarrier band, this approach leads to self-cancellation of the intrinsic interference and hence, it improves robustness to frequency selective channels.

This work is inclined toward more practical scenarios than what is available in the literature and considers the systems with imperfect CSI, too. The focus of this paper is on downlink transmission while using the channel estimate in the uplink for precoding. Channel estimation in massive MIMO is normally limited to the uplink. 
Real-field orthogonality in FBMC makes channel estimation a more complex task than in OFDM. 
The majority of proposed FBMC channel estimation techniques in literature are based on the interference approximation method (IAM), e.g., see \cite{lele2008channel}. IAM, as a frequency domain method, requires that the channel delay spread be much smaller than the symbol interval to satisfy a flat channel response per subcarrier band.  A review of the IAM-based channel estimation methods for FBMC is provided in \cite{kofidis2013preamble2}. Time domain channel estimation was proposed to address issues in IAM-based channel estimation \cite{caus2012transmitter,kofidis2013preamble,kofidis2014short,kong2014time,kofidis2017preamble,zhang2017iterative,kofidis2015preamble,singh2019time,Hoss2006:Spectrally}. Among these methods, the ones in \cite{kofidis2015preamble,Hoss2006:Spectrally} are more suitable for multi-user scenarios where they do not require guard symbols between different users pilot signals. Another interesting approach that has recently emerged in the literature is based on the idea of deploying a superimposed preamble with the data symbols, \cite{kofidis:hal-02889973,chen2017superimposed}.
In this approach, no isolation between the preamble and data symbols is required. While superimposed pilots improve spectral efficiency, they require iterative interference cancellation. This process purifies the training signal from intrinsic interference that is caused by  the data symbols at the expense of an increased computational load. 

In this paper, we employ the same pilot structure and time domain channel estimation method as the one in \cite{hosseiny2021fbmc}. \textcolor{black}{This approach estimates the channel in the time domain and utilizes the minimum number of pilots, equivalent to the length of the channel impulse response per user. By exploiting intrinsic interference, joint estimation of the channel impulse response for users eliminates the need for guard symbols between their pilot signals. These actions enhance the spectral efficiency of the system and improve performance in delay-stringent networks.}

\color{black}
Imperfect CSI leads to performance degradation, and its effect on the uplink of FBMC-based massive MIMO was studied in \cite{hosseiny2021fbmc}. The design of an effective precoder in presence of channel estimation errors has been the subject of multiple publications on \color{black}  narrow-band \cite{israr2017performance,lu2019robust,choi2020joint,qiu2018downlink} and  OFDM-based \cite{dinh2020pca,colon2015linear} massive MIMO systems.  
Authors in 
\cite{israr2017performance} evaluated the performance of narrow-band Massive MIMO with linear precoding techniques. Spectral efficiency, represented by achievable rate, and energy efficiency for zero forcing (ZF) and maximum ratio transmission (MRT) precoders, under imperfect CSI, are investigated in this work. 
In \cite{lu2019robust}, the statistical CSI for each user equipment (UE) at the base station (BS) is characterized under a jointly correlated channel model, accounting for channel estimation error, channel aging, and spatial correlation. The proposed precoding algorithm in \cite{lu2019robust} reaches a stationary point of the expected weighted sum-rate maximization problem. Considering imperfect CSI, an optimization framework with a unified solution for joint user selection, power allocation, and precoding of multi-cell massive MIMO is introduced in \cite{choi2020joint}. This solution has a superior performance to linear precoders.
\color{black}
The impact of mutual interference between two types of imperfection, i.e., statistical CSI and imperfect instantaneous CSI, is analyzed in \cite{qiu2018downlink} where the authors propose modified ZF and minimum mean square error (MMSE)  precoders. 
\color{black}
A linear precoding approach for a sub-band of OFDM-based massive MIMO systems that combines the conventional linear precoders with the principal component analysis technique is proposed in \cite{dinh2020pca}. The influence of imperfect CSI and hardware impairments on the downlink of OFDM-based massive MIMO is investigated in \cite{colon2015linear}. 

Although it is straightforward to extend narrow-band analysis to OFDM-based systems, extending the analysis to FBMC-based massive MIMO requires careful consideration and thorough investigation. Publications on the precoder design for FBMC-based systems with imperfect CSI are limited to MIMO systems \cite{ruyet2014precoding,doanh2021combining}. These works analyze FBMC-based MIMO performance in presence of imperfect CSI, considering perfect channel reciprocity.  It is worth noting that to the best of our knowledge, there is no work on the precoder design for FBMC-based massive MIMO in presence of frequency selective channels, imperfect CSI, and reciprocity calibration errors. 
\color{black}

As of today, most publications on FBMC-based massive MIMO have focused on the uplink, \cite{aminjavaheri2018filter,aminjavaheri2017prototype,Hoss2006:Spectrally,hosseiny2021fbmc,singh2019time}. A few of these publications have assumed perfect reciprocity and, accordingly, have noted the proposed uplink detection methods may be reversed to design precoders for the downlink of the same link, e.g., see \cite{aminjavaheri2018filter}.  However, the assumption of perfect reciprocity may not be valid, both due to channel aging in time division duplexing (TDD) and the differences in radio chains (even after calibration) in the uplink and downlink directions.  Works such as \textcolor{black}{\cite{chopra2020blind,Shahabi2019novel,raeesi2018performance,mi2017massive} } have investigated imperfect reciprocity problem for narrow-band systems. \color{black}
\color{black} 
Authors in \cite{mi2017massive} present an analysis of the effects of reciprocity calibration and channel estimation errors on conventional linear precoders in a TDD massive MIMO system, while also considering channel estimation errors. The closed-form expressions for the output SINR are derived for MRT and ZF precoders. \cite{Shahabi2019novel} also explores the problem of reciprocity calibration in a massive MIMO system and design precoders capable of compensating for non-reciprocal channels and imperfect CSI. Authors in \cite{raeesi2018performance} examine the efficiency of the downlink of TDD-based massive MIMO system with linear precoders, considering the combined effects of channel non-reciprocity and imperfect CSI. \color{black}
This study showed that the effect of imperfection on precoders performance could be severe, leading to saturation of large antenna effects.
\color{black}
While the extension of \textcolor{black}{narrow band annalysis} to OFDM-based systems is straightforward, it requires particular attention and investigation to be extended to FBMC-based massive MIMO.

\color{black}

{\color{black}In FBMC, the challenges faced in the uplink, \cite{aminjavaheri2018filter,hosseiny2021fbmc,singh2019uplink}, such as the inadequacy of single-tap equalization and the need for subcarrier flattening, also apply to the downlink scenario. However, there is a lack of research addressing these issues. Furthermore, similar to single carrier and OFDM systems, here also, perfect channel reciprocity between the uplink and downlink may not be attainable due to hardware limitations and calibration errors. These imperfections in reciprocity lead to performance degradation in the downlink of FBMC  massive MIMO systems. Hence, it is crucial to investigate these effects and develop techniques to mitigate their impact and enhance the overall performance of such systems in practical deployments.} In \cite{Hosseiny2022downlink}, as an initial work on this topic, we laid down the foundations for downlink precoding of FBMC  massive MIMO systems. \color{black} In particular, we considered channel estimation and reciprocity calibration errors for precoder design in the asymptotic regime. However, our analysis was limited to co-located antenna deployments. 
Hence, in this paper,  we extend the results of \cite{Hosseiny2022downlink} to massive MIMO setups with distributed antennas, also known as \textit{cell-free} massive MIMO.

\color{black} We also propose precoding techniques without the assumption of frequency-flat channels over the subcarrier bands in presence of imperfect CSI and channel reciprocity calibration errors. To this end, we formulate the precoding problem in terms of an equivalent downlink channel while taking into account the presence of the aforementioned imperfections. Our analytical derivations reveal that the CSI and channel reciprocity imperfections converge to the statistics of these errors. This paves the way towards the design of a fractionally spaced prefilter (FSP) that takes into account imperfection correction procedure prior to precoding. At the precoding stage, any of the conventional linear precoding methods may be deployed.

We take note that the imperfection statistics are not always available at the BS. Thus, for such cases, we propose a downlink training procedure to find and compensate the residual effects after signal reception at the UE. This procedure comes at the expense of slight performance loss when compared to the case where the imperfection statistics are available at the BS. Since, in cell-free massive MIMO architecture, the antennas are distributed in space, different large-scale fading coefficients affect the received signal at each UE. Hence, power allocation is necessary in the downlink of cell-free massive MIMO to balance the trade-off between fairness and average signal-to-interference-plus-noise ratio (SINR). This leads us to investigate fractional power allocation in  the FBMC-based cell-free massive MIMO architecture.
\color{black}
Furthermore, an antenna selection method for the cell-free massive MIMO based on the received SINR in the uplink direction is proposed. The access point (AP) selection forms a user-centric architecture in the cell-free setup that limits the AP-UE connection to an optimum set. Finally, the simulation results that evaluate the performance of the proposed downlink precoding and the corresponding correction methods are presented.

\color{black} To summarize, the main contributions of this paper are the following: (1) We formulate the precoding problem in the downlink of FBMC-based massive MIMO by assuming a perfect reciprocal channel. (2) We propose a two-stage precoder structure whose first stage involves a short fractionally spaced prefilter (FSP) at each subcarrier for flattening the equivalent channel. This stage is then followed by a conventional linear precoder. (3) In our proposed FSP design, we take into account imperfect CSI and calibration reciprocity error effects and analytically derive their effect on the received signal.  (4) We formulate the downlink of FBMC-based cell-free massive MIMO for the first time and investigate power allocation for this setup. (5) We propose an access point selection procedure for the cell-free setup based on the received SINR of the uplink. (6) We examine the effect of imperfect CSI and reciprocity calibration error on the cell-free massive MIMO. We show that these effects converge to the statistical characteristic of these parameters and propose modifications.

\color{black}

\color{black}

The rest of the paper is organized as follows. Section~\ref{sec:flatfbmc} presents principles of FBMC in the downlink of massive MIMO, assuming a flat response at each subcarrier. In Section~\ref{sec:selective_precoding}, we propose a two-stage precoding to overcome frequency selectivity in the channel. The CSI and channel reciprocity errors and their impact on the proposed precoding method are studied in Section~\ref{sec:calibration}. We also propose compensation methods for relaxing the effects of both errors.  Section \ref{sec:simulation} presents numerical results, corroborating our theoretical studies. Finally, the paper is concluded in Section \ref{sec:conclusion}.

\vspace{3mm}
\noindent\textit{Notations:} Matrices, vectors and scalar quantities are denoted by boldface uppercase, boldface lowercase and normal letters, respectively.  $A(m,l)$ represents the element in the $m^{\rm th}$
row and the $l^{\rm th}$ column of $\bA$ and $\bA ^{-1}$ signifies the inverse of $\bA$. $\bI_M$ is the identity matrix of size $M \times M$. Superscripts $(\cdot)^{-1}$, $(\cdot)^{\rm T}$, $(\cdot)^{\rm H}$ and $(\cdot)^*$ indicate inverse, transpose, conjugate transpose, and conjugate operations, respectively. $\mathfrak{R}\{\cdot\}$,  $\mathds{E}\{\cdot\}$, ($\downarrow M$) and $\star$  represent real value, expectation, $M$ fold decimation, and linear convolution operators, respectively. 
Finally, $\delta_{\textcolor{black}{ki}}$ represents the Kronecker delta function.

 \vspace{-1mm}
\section{Downlink FBMC System Model}
\label{sec:flatfbmc}

In \textcolor{black}{FBMC-OQAM}, real-valued data symbols are placed on a regular time-frequency grid with the time and frequency spacings of $T/2$ and $1/T$, respectively. Each data symbol on the grid has a $\pm\frac{\pi}{2}$ phase difference with its neighbours. This is to avoid interference between the data symbols and hence make them orthogonal in the real domain. The data symbols are pulse-shaped with a prototype filter $f[l]$, where $f[l]$ is designed such that $q[l]=f[l]\star f^*[-l]$ satisfies the Nyquist criterion. Therefore, assuming $M$ number of subcarriers, the Nyquist pulse $q[l]$ has zero crossings every $M$ samples. Considering a narrow bandwidth for each subcarrier such that the data symbols experience approximately flat fading channels, per subcarrier precoding can be deployed in the downlink \cite{marzetta2016fundamentals}.

Let us consider a single-cell massive MIMO setup including a BS equipped with $N$ antennas and $K$ single-antenna UEs\textcolor{black}{\footnote{\textcolor{black}{This setup can be easily extended to multiple antenna users by treating each user antenna as if it is a single-antenna user in our scenario.}}}. \color{black} Let $d^k_{m,n}$ be the real-valued data symbol of user $k$ at the subcarrier $m$ and the time slot $n$. For each frequency-time instant $(m,n)$, the precoder collates  data symbols  $d_{m,n}^k, k=1,2,\cdots,K$ for all users and forms the transmit signal vector. 
\color{black}
\be 
s^i_{m,n} = \sum_{k=0}^{K-1}\sqrt{q_{k}}P^{i,k}_m d^{k}_{m,n},\label{eq:s_{m,n}}
\ee
\color{black}
\textcolor{black}{where $q_k$ is the allocated power to user $k$ and} $\bP_m$ is the $N \times K$ precoding matrix {\color{black}with elements $P^{i,k}_m$,} for $k=0,\ldots,K-1$ and $i=0,\ldots,N-1$. Two choices of $\bP_m$ have been introduced in the literature: (i) maximum ratio transmission (MRT), and (ii) zero-forcing (ZF), expressed as, \cite{nam2016new,rusek2012scaling},
 \be 
 \bP_m = \begin{cases}
 \bH_m^{\rm H} \bD_m^{-1}, & \text{for MRT,}\\
 \bH_m^{\rm H} \big(\bH_m \bH_m^{\rm H}\big)^{-1}, & \text{for ZF.}\\
\end{cases}
\label{eq:combiners}
\ee
\color{black}
Here, $\bH_m$ is the $K\times N$ channel matrix with elements $H_m({k,i})$ representing the channel gains between UE $k$ and BS antenna $i$ at the center of subcarrier $m$, i.e.,  $H_m({{k,i}}) \triangleq \sum_{l=0}^{L-1} h_{{k,i}}[l] e^{-j\frac{2 \pi m l}{M}}$, {\color{black}where $h_{k,i}[l]$ is the respective channel impulse response \textcolor{black}{with length $L$.}} In MRT, the $K \times K $ diagonal matrix $\bD_m$, with the diagonal elements $D_m^{k,k} = \sum _{i=0}^{N-1}|H_m({{k,i}})|^2$, normalizes the precoder output. Assuming reciprocal channels in the uplink and downlink, the estimated channel responses in the uplink phase are used for precoding. 

\color{black}

While MRT precoder relies on the large size of $N$ to suppress interference among different user, ZF precoder removes interference completely for any choice of $N\ge K$, and known to be the best linear decoder; e.g., see \cite{liang2014low}. Taking note of this, all of our results in this paper are based on ZF precoder. 

\color{black}

After precoding, the transmit signal at the BS antenna $i$ can be formed by passing the symbols $s^i_{m,n}$ through the synthesis filterbank (SFB)
\begin{equation}
x_i[l] = \sum_{m=0}^{M-1} \sum_{n=-\infty}^{ \infty} s^i_{m,n} f_{m,n}[l] ,
\label{bbeq}
\end{equation}
where
$f_{m,n}[l] = f\big[l-n\frac{M}{2}\big] e^{j2\pi ml/M}e^{j\pi (m+n)/2}$
is the modulated, time shifted, and phase-adjusted pulse-shape that carries $s^i_{m,n}$. Finally, the received signal at user $k$ can be obtained as
\be
r_k[l] = \sum_{i=0}^{N-1} x_i[l] \star h_{k,i}[l] + \eta_k[l],
\label{eq:rec_antenna}
\ee
where $\eta_k[l] \sim \mathcal{CN}(0,\sigma^2_{\eta})$ is the additive white Gaussian channel noise at the UE $k$. 
{\color{black} Considering co-located BS antennas, we can assume the same power-delay profile (PDP) between the BS antennas and any given user $k$. This PDP is denoted by  $p_{k}[l]$ for $l=0,\ldots,L-1$. 
\color{black} The large-scale fading coefficient, $\beta_{k,i}$, that captures the shadowing effect depends on the distance between each user $k$ and BS antenna $i$ \cite{ngo2017cell,nayebi2015cell}. In co-located massive MIMO, the channel between a given user $k$ and BS antenna $i$ is modeled with the  PDP, $p_{k,i}[l]$ where $\sum_{\color{black} l} p_{k,i}[l] =\beta_{k,i} $. As the BS antennas are co-located, large-scale and shadow fading is the same for all of them, hence, we drop the BS antenna index from the PDP and large-sale fading coefficient for massive MIMO with co-located antennas. Accordingly, for user $k$, the channel taps are independent of one another and their distribution follows $\mathcal{CN}(0,p_{k}[l])$. \color{black}

Assuming perfect synchronization and the availability of perfect channel knowledge at the BS, the data symbols of each user can be extracted as
\be
\hat{d}_{m,n}^k = \Re \{(r_k[l] \star \color{black} f_{m,n}[l]\color{black})|_{{l=\frac{nM}{2}}} \}.
\label{d^hat}
\ee
In FBMC, the assumption of flat fading subcarrier channels can never be satisfied no matter how narrow the subcarrier bands are made. Furthermore, to avoid spectral efficiency loss, it is always desirable to keep the subcarrier bands as wide as possible; \textcolor{black}{see \cite{farhang2014} and \cite{farhang2014massive} } for some explanations along this line. Moreover, we may recall from \cite{farhang2014} that channel hardening effect in massive MIMO systems allows one to widen the subcarrier bands. 
However, investigations in \cite{aminjavaheri2018filter} have revealed that  the hardening effect flattens the channel to a limit. Hence, an additional equalization/precoding step is required for (near) perfect flattening of the channel over each subcarrier band. Moreover, imperfections in the available CSI that may originate from channel aging, estimation error, as well as reciprocity calibration errors have to be compensated. To tackle these practical problems, in Section~\ref{sec:selective_precoding}, we propose a prefilter that flattens each subcarrier channel, and in  Section~\ref{sec:calibration}, we address the imperfect CSI and channel reciprocity issues.

\section{Co-Located Massive MIMO Precoding} 
\label{sec:selective_precoding}

Using \eqref{eq:rec_antenna}, \eqref{d^hat} can be expanded as
\begin{align}\label{eq:dhatmnk1}
\hat{d}_{m,n}^k \!= &\Re \bigg\{ \!\sum_{i=0}^{N-1} \! \big(x_i[l] \star h_{k,i}[l] \star \color{black}f_{m,n}[l]\color{black}\big)\big|_{l=\frac{nM}{2}} + \eta_{m,n}^k \bigg\},
\end{align}
where $\eta_{m,n}^k$ represents the noise effect after filtering, sampling, and phase adjustment. Recalling \eqref{eq:s_{m,n}} and \eqref{bbeq}, \eqref{eq:dhatmnk1} can be simplified as
\begin{align}
&\hat{d}_{m,n}^k =   \Re \bigg\{ \sum_{m',n',k'}   d_{m',n'}^{k'} g_{m,m'}^{k,k'}[n-n']  +\eta_{m,n}^k \bigg\},
\label{eq:eqxpandedEstimate}
\end{align}
where
\begin{align}\label{eq:g_{mm'}^{k,k'}[n]}
 g_{mm'}^{k,k'}[n] = \big(f_{m'}[l] \star h_{k,k',{m\textcolor{black}{'}}}^{\rm (eqvlt)}[l] \star f_m^*[l]  \big)\big|_{l=\frac{nM}{2}},
\end{align}
\color{black}
\be\label{eq:equivalent-channel-general}
h_{k,k',{m\textcolor{black}{'}}}^{\rm (eqvlt)}[l]=\sum_{i=0}^{N-1} \textcolor{black}{\sqrt{q_{k\textcolor{black}{'}}}}(P_{m\textcolor{black}{'}}^{{i,k'}})^*h_{\textcolor{black}{k,i}}[l],
\ee
\color{black}  Equation \eqref{eq:eqxpandedEstimate} includes the effects of transmit filtering, precoding, the multipath channel, and the receive filtering.} 
{\color{black} The estimate of the symbol is guaranteed when a flat gain over each subcarrier is assumed.} 
As shown in \cite{ngo2013energy}, for a large number of antennas, all the above precoders converge to \color{black}${\rm diag}([N\beta_{0}, ..., N\beta_{K-1}])^{-1} \bH_m$. \color{black}Accordingly, the equivalent channel over the subcarrier band $m$ between the user $k$ and the precoder input intended for user $k'$ may be expressed as 
\be\label{eq:equivalent-channel}
h_{k,k',m}^{\rm (eqvlt)}[l]=\frac{\textcolor{black}{\sqrt{q_{k'}}}}{N \textcolor{black}{\beta_k'}}\sum_{i=0}^{N-1}(H_m({{k',i}}))^*h_{\textcolor{black}{k,i}}[l].
\ee 
Using the law of large numbers, as the number of the BS antennas grows large,  $h_{k,k',m}^{\rm (eqvlt)}[l]$ vanishes to zero, when $k\ne k'$. Additionally, when $k=k'$, it can be shown that $h_{k,k,m}^{ \rm (eqvlt)}[l]$ in {\color{black}{\eqref{eq:equivalent-channel}}} converges to \cite{aminjavaheri2018filter,hosseiny2021fbmc}
\be\label{eq:pmk[l]}
\bar{p}_{m,k}[l]= \frac{\textcolor{black}{\sqrt{q_{k}}}}{ \textcolor{black}{\beta_k}} p_{k}[l]   e^{j2\pi lm/M},
\ee
where $p_{k}[l]$ is the channel PDP between the user terminal $k$ and the BS antennas.
 The above equation shows the residual channel that breaks the Nyquist property is characterized by the PDP of the channel. This effect can be pre-compensated at each subcarrier before precoding.  Hence, we propose a FSP, that, for any $m$, covers the $m$-th subcarrier band, including the overlapping parts of the band with the adjacent subcarriers. {\color{black} Further details on the design of an FSP can be found in \cite{farhang2008signal,farhang2013adaptive}.} This prefiltering also eliminates the intrinsic interference from the adjacent bands. The prefilter design may be a ZF or an MMSE one that can provide a satisfactory performance with minimum number of taps. Details of such filter designs are explained in \cite{1457566} and \cite{hosseiny2021fbmc}. {\color{black}From \eqref{eq:pmk[l]}, one may realize that the prefilter for each subcarrier relates to the frequency-shifted version of the PDP. This implies that the prefilter is the frequency-shifted version of the base-band prefilter and needs to be calculated only once for each user.} As it will be shown in Section~\ref{sec:simulation}, the proposed prefiltering significantly improves the output SINR by flattening the channel over each subcarrier band.

In practical systems where the number of BS antennas is not large enough, the formulated FSP may not be adequately effective. This is because the equivalent channel does not converge to the frequency shifted PDP in \eqref{eq:pmk[l]}. In this case, using \eqref{eq:equivalent-channel-general} and assuming perfectly compensated multiuser interference at the precoding stage, the equivalent channel response for $k=k'$ is $h_{k,k,m}^{\rm (eqvlt)}[l]=\sum_{i=0}^{N-1}\textcolor{black}{\sqrt{q_{k}}}(P_m^{{i,k}})^*h_{\textcolor{black}{k,i}}[l]$. Consequently, the FSP needs to be designed based on this equivalent channel. It is worth noting that this operation is performed before precoding at the BS to avoid increasing the user equipment complexity. 

\begin{figure*}[ht]
		\centering
		\includegraphics[scale=0.38,trim={0 0 0 0},clip]{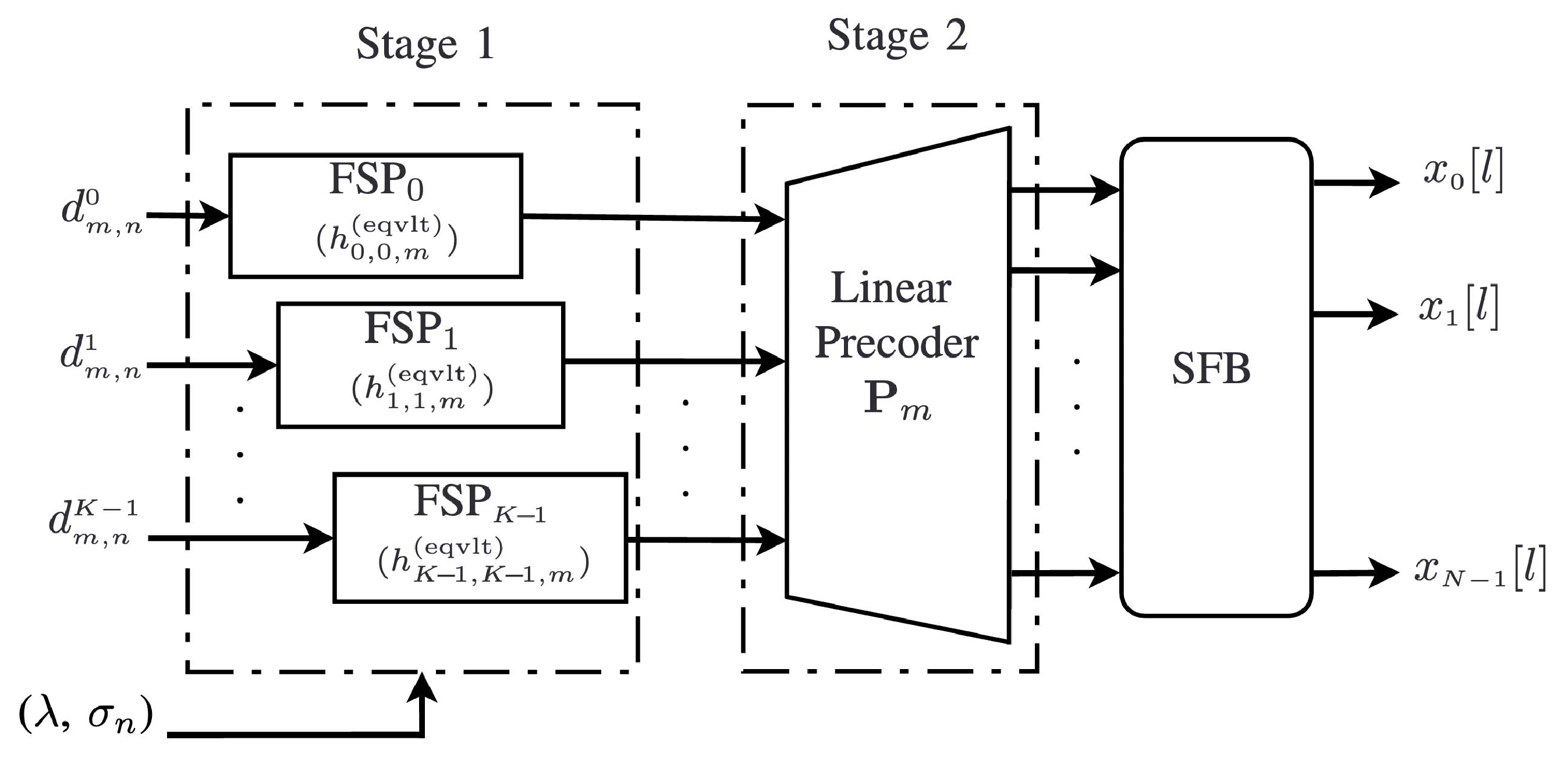}
		\caption{The proposed two-stage precoding scheme. The first stage is a set of FSPs that flattens the channel for the second stage of conventional linear precoding. Conventional linear precoding repeats for each subcarrier $m$ and every user $k$ to remove ISI and ICI.
		}
		\label{blockdiagram_DL}
	\end{figure*}

Our proposed two-stage prefiltering and downlink precoder structure is illustrated in Fig.~\ref{blockdiagram_DL}. In this structure, the data symbols are passed through a set of FSPs followed by a  conventional linear precoder at each subcarrier $m$ and for every user $k$ for removing intersymbol interference (ISI) and  intercarrier interference (ICI). Each FSP is designed based on the indicated equivalent channel, following \cite{hosseiny2021fbmc}. The first stage can be thought of as a channel flattening step, which makes it possible for the single-tap precoder per subcarrier to perform optimally. Simulation results that confirm the efficacy of this prefilter design are provided in Section~\ref{sec:simulation}.

\section{Cell-Free Massive MIMO Precoding} 
\label{sec:cell_free}
Cell-free massive MIMO has recently emerged as an attractive architecture for future wireless networks. Most of the available literature on this topic are based on OFDM or narrowband communication systems. FBMC-based cell-free massive MIMO was first introduced in \cite{hosseiny2021fbmc}, for the uplink scenario. In this section, we cast the downlink scenario into a mathematical formulation. We also present an antenna selection method for eliminating the antennas whose contribution to the received signal at each UE is negligible.

%

\subsection{Cell-free/Distributed Antenna Downlink System Model}
\label{sec:system_model}
We consider a case where there are  $N_{\rm AP}$ distributed APs, equipped with $Q$ antennas each. As in \cite{bjornson2019making}, a central processing unit (CPU) transmits the precoded symbols to the total of $N=QN_{\rm AP}$ antennas through a backhaul link. It is worth mentioning that a given antenna is only assigned to one AP, i.e., the antenna indices of different APs belong to mutually disjoint subsets of the available antennas. 

The large-scale fading coefficient, $\beta_{\color{black} k,i}$, depends on the distance between each user $k$ and antenna $i$ and captures the shadowing effect \cite{ngo2017cell,nayebi2015cell}. In cell-free massive MIMO, the channel between a given user $k$ and antenna $i$ is modeled as multiplication of the associated large-scale fading and multipath channel with a normalized PDP, i.e., \textcolor{black}{$\sum_l p_{\color{black} k,i}[l] =  \beta_{\color{black} k,i}$}. It is worth noting that large-scale fading coefficients for a given user $k$ and all the antennas of a given AP, \textcolor{black}{$N_{\rm AP}$}, are equal.

Similar to the uplink scenario, which requires power control to provide fairness among different users  \cite{ngo2017cell}, the variations in large-scale fading coefficients necessitate power allocation for the downlink transmission. As shown in \cite{hosseiny2021fbmc}, fractional power control provides a similar performance to OFDM for the FBMC-based systems. This process involves increasing the probability of getting an SINR close to the average SINR for all the users, i.e., holds fairness. To address the power allocation problem, we employ the fractional power allocation proposed for the OFDM-based cell-free massive MIMO in \cite{nikbakht2020uplink}. Accordingly, here, we set the amount of power dedicated to a given user $k$ by antenna $i$ in the FBMC-based cell-free massive MIMO as
   \be
q_{\color{black} k,i} \propto \frac{\beta_{\color{black} k,i}}{(\sum_{i'=0}^{N-1}\beta_{\color{black}k,i'})^{\nu} \big(\sum_{k'=0}^{K-1}\frac{\beta_{\color{black}k',i}}{(\sum_{i'=0}^{N-1}\beta_{\color{black}k',i'})^{\nu}}\big)^{\gamma}}.\label{p_ik}
\ee
In the fractional power allocation, two parameters $\nu$ and $\gamma$ are used to adjust the power and strike a balance between fairness and average SINR. By design, $\nu \in [0.5,\; 0.7]$ and $\gamma \in [0.8,\; 1.4]$ are the recommended ranges, \cite{nikbakht2020uplink}. 
In \eqref{p_ik},  $\nu$ adjusts the amount of power that a specific user receives from antenna $i$ while $\gamma$ adjusts the power with respect to the total power transmitted by a specific antenna. 


By using the formulation developed for uplink cell-free in \cite{hosseiny2021fbmc} and the downlink formulation in Section \ref{sec:flatfbmc}, the transmit signal of antenna $i$ can be expressed as
\color{black}

\color{black}
\be
x_i[l] = \sum_{m=0}^{M-1} \sum_{n=-\infty}^{ \infty} \sum_{k=1}^{K} \sqrt{q_{\color{black} k,i}} 
(P_m^{\textcolor{black}{\color{black} k,i}}) ^* d_{m,n}^{ k} f_{m,n}[l].
\label{bbeq_cell_free}
\ee
\color{black}
Note that the precoders in equation (2) are applicable to cell-free case as a total number of  $N$ antennas are among all APs.

\color{black}
The precoded symbols in the cell-free scenario are assumed to be obtained from \eqref{eq:s_{m,n}} using a linear precoder from \eqref{eq:combiners}. 
By expanding \eqref{bbeq_cell_free} and using \eqref{eq:rec_antenna} and \eqref{eq:s_{m,n}}, we obtain the received symbol of user $k$ at the time-frequency bin $(m,n)$ as
\begin{align}
&\hat{d}_{m,n}^k =   \Re \bigg\{ \sum_{m',n',k'}   d_{m',n'}^{k'} g_{m,m'}^{k,k'}[n-n']  +\eta_{m,n}^k \bigg\},
\label{eq:eqxpandedEstimate_cell}
\end{align}
where
\begin{align}\label{eq:g_{mm'}^{k,k'}[n]_cell}
 g_{mm'}^{k,k'}[n] = \big(f_{m'}[l] \star h_{k,k',m}^{\rm (eqvlt)}[l] \star f_m^*[l]  \big)\big|_{l=\frac{nM}{2}},
\end{align}
\be\label{eq:equivalent-channel-cellfree}
h_{k,k',{m\textcolor{black}{'}}}^{\rm (eqvlt)}[l]=\sum_{i=0}^{N-1} \sqrt{q_{k\textcolor{black}{'},i}}(P_{m\textcolor{black}{'}}^{k',i})^*h_{\color{black} k,i}[l],
\ee
and $P_{m\textcolor{black}{'}}^{\textcolor{black}{\color{black} k',i}}$ is the precoder coefficient given by \eqref{eq:combiners}. The equivalent channel in (\ref{eq:equivalent-channel-cellfree}) after precoding causes SINR degradation and limits the achievable rate. As noted earlier, in FBMC-based massive MIMO with co-located antennas, the equivalent channel converges to the PDP of the underlying channel. In contrast, in cell-free massive MIMO, the channel PDP between each UE and different APs varies significantly. Thus, the equivalent channel does not converge to a particular PDP or an equivalent PDP. Considering these observations, we propose a per-subcarrier per-user FSP based on \eqref{eq:equivalent-channel-cellfree}.

\subsection{AP Selection}
\label{sec:cell_fee_AP_selection}
In a canonical cell-free massive MIMO architecture where all the APs serve all the users \cite{buzzi2018user}, the contribution of some APs to the received power at each UE may be negligible. The relative variations in large-scale fading between a given UE and different APs result in relatively small received power from some APs. Hence, the user-centric approach, \cite{ngo2017total}, where only a subset of APs with higher contributions to the received signal power at a given UE are selected, is more effective. This is of a paramount importance as  APs have a limited power budget. Furthermore, efficient allocation of the APs to the users reduces the backhaul traffic and improves the overall efficiency of the wireless networks. Apart from all the benefits, the process of selecting the optimal subset of APs for each UE, i.e., antenna selection, is a crucial part of this approach. In \cite{ngo2017total}, two antenna selection methods for the downlink of cell-free massive MIMO architecture are introduced. These are based on the received power and large-scale fading. \textcolor{black}{ It is worth noting that an antenna selection in the uplink can be performed with similar goals to the downlink. Therefore, by limiting antenna selection to one direction, the complexity and overhead of TDD communication reduces.} In this paper, we propose an antenna selection technique that starts from the uplink phase. 

In the uplink, our proposed antenna selection technique helps to alleviate the effect of noise and interference at the signal-combining stage that is performed by the CPU. In this technique, for signal detection, we propose to only combine the received signals at a subset of APs whose signal-to-noise ratios (SNR)s for a given user, $k$, are above a certain threshold. Therefore, for each user, $k$, we form a set, $\mathcal{B}_k$, that includes the antenna indices whose SNRs are above the set threshold. \textcolor{black}{It is worth noting that the value of SNR can be measured at the uplink channel estimation phase by using each users pilots.} Thanks to channel reciprocity in the TDD mode, we propose to deploy the same subsets of antennas that were used for combining and detection of different users' signals in the uplink for antenna selection in downlink transmission.  

\begin{figure}
		\centering
  \hspace{-12.5mm}
		\includegraphics[scale=0.33,trim={0 0 0 0},clip]{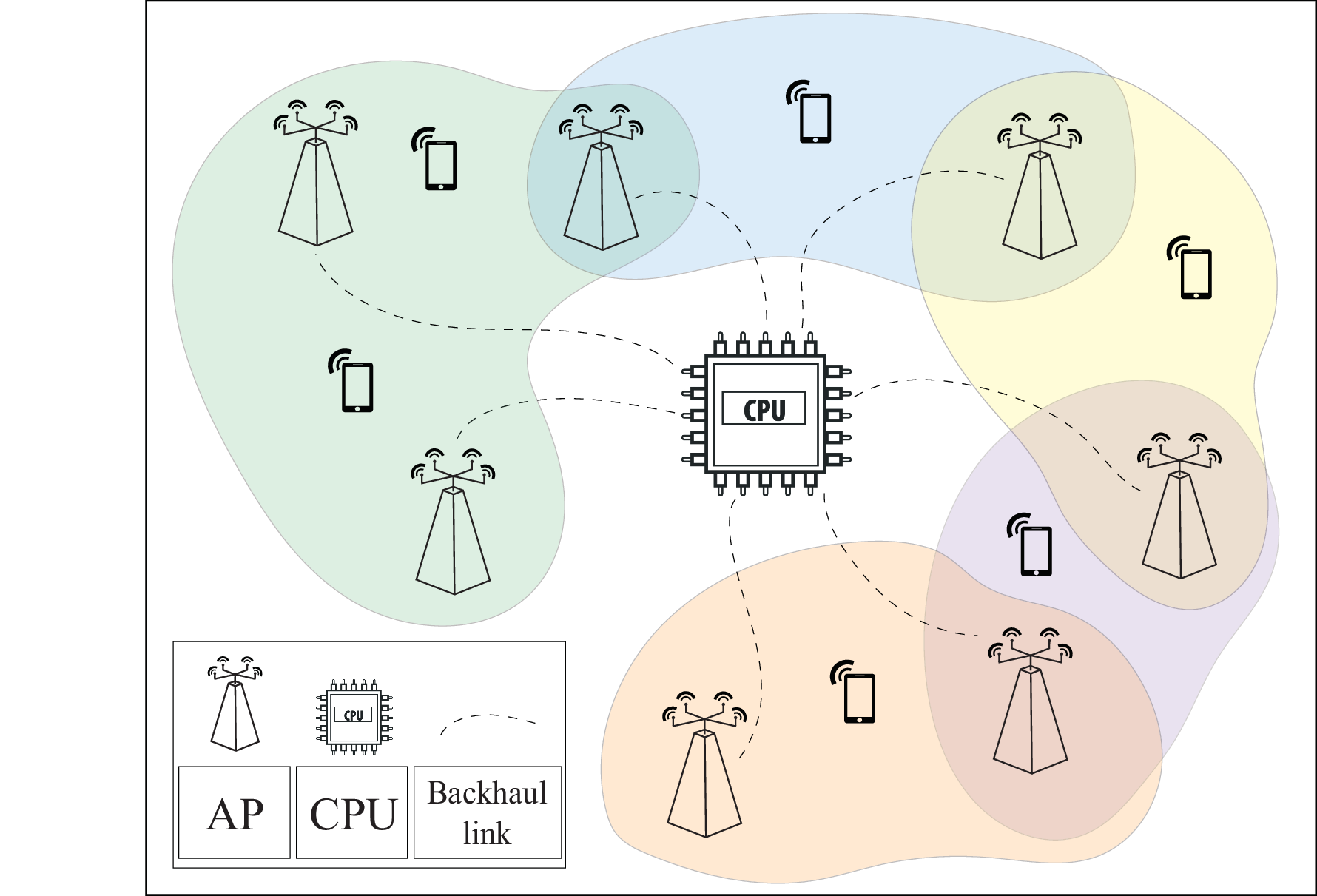}
	\caption{Cell-free massive MIMO network architecture with different colors illustrating the regions of APs serving each user.}
		\label{fig:cell_free_diagram}
		\vspace{-0.4 cm}
	\end{figure}

In Section~\ref{sec:simulation}, we numerically evaluate the performance of our proposed antenna selection technique where we show its effective performance under imperfect reciprocity. An example of a cell-free network is depicted in \textcolor{black}{Fig.} \ref{fig:cell_free_diagram} where the areas of different users' antenna subsets are highlighted by different colors.

\color{black}

 \color{black}
\section{Precoding with Imperfect Reciprocity}
\label{sec:calibration}
In the developments so far, we considered perfect channel knowledge at the BS and perfect channel reciprocity in both uplink and downlink directions. However, these assumptions may not be accurate in practical systems, as the channel estimates at the BS may suffer from estimation errors.
Moreover, while the uplink and downlink propagation channels may be the same, different transceiver chains at the UEs and the BS break the channel reciprocity. Channel aging is another effect that may lead to channel reciprocity error.
Hence, imperfect channels should be considered in designing downlink precoders. 

In the past, a number of channel reciprocity calibration techniques were developed \cite{kaltenberger2010relative}. However, reciprocity calibration error can still lead to detrimental effects on the downlink transmission \cite{mi2017massive}. Recently, considering the channel estimation technique in \cite{Hoss2006:Spectrally} and the associated channel estimation error statistics that are provided in \cite{hosseiny2021fbmc}, we have studied the effects of inaccurate channel estimates on signal detection.  Based on the results of \cite{hosseiny2021fbmc}, the channel estimation error for the tap $l$ between BS antenna $i$ and user $k$, $\Delta h_{\color{black} k,i}[l]$, can be approximated as a zero mean complex Gaussian random variable with the variance $\sigma_{\rm et}^2 =  \frac{\rm MSE}{K \times L}$. The MSE of the estimation method is also calculated in \cite{hosseiny2021fbmc}. It  has also been noted that the estimation error at a given subcarrier $m$ follows complex Gaussian distribution with the variance $\sigma_{\rm ef}^2=L \sigma_{\rm et}^2.$
That is, $\Delta H_{m}({\color{black} k,i}) \sim \mathcal{CN}(0,\sigma^2_{\rm ef})$. It is worth noting here that the subscripts `et' and `ef' refer to estimation errors in time and frequency domains, respectively. 

In the following subsections, we use the channel reciprocity calibration error model to analyze the imperfect CSI and calibration error effects on the precoder output in large and moderate antenna regimes for co-located massive MIMO. This paves the way towards our proposed precoder designs that pre-compensate these errors for both co-located and cell-free massive MIMO.

\color{black}
\subsection{Reciprocity Calibration Error Model}
\textcolor{black}{Here, we consider a setup with each antenna connected to an independent radio frequency (RF) chain. Furthermore, we assume channel estimation and downlink transmission being performed during the coherence time of the channel with a negligible antenna coupling effect. \textcolor{black}{Assuming the reciprocity calibration methods in \cite{kaltenberger2010relative}, RF chains cannot be considered to be perfectly matched. Hence, a residual calibration error always remains.} Reciprocity error can be considered as a linear function of time, power, and temperature \cite{3GPPmeasurement}. These dependencies are measured during the calibration process. Moreover, the non-linear effects cause reciprocity calibration errors and are modeled as random variables. The mismatch effect, similar to the calibration effect, can be modeled as an independent complex gain for each RF chain \cite{3GPPmeasurement,mi2017massive,chopra2020blind,kaltenberger2010relative}. Accordingly, the uplink and downlink channels, with reciprocity calibration error, can be modeled as \cite{kaltenberger2010relative}
\be \label{eq:imperfect_reciprocty_NB}
H_m^{\rm u}(k,i) = \xi_{{\rm r},i}^{m} e^{j\phi_{{\rm r},i}^m} H_m(k,i),
\ee
and
\be \label{eq:imperfect_reciprocty}
H_m^{\rm d}(k,i) = \xi_{{\rm t},i}^{m} e^{j\phi_{{\rm t},i}^m} H_m(k,i), 
\ee
where $\xi_{{\rm t},i}^{m}$ and $\phi_{{\rm t},i}^m$ denote the magnitude and phase of the calibration error for subcarrier $m$, respectively, and $\xi_{{\rm r},i}^m$ and $\phi_{{\rm r},i}^m$ represent the equivalent variables of the receive RF chain for antenna $i$ and subcarrier $m$.
The statistical characteristics of these errors can be obtained using empirical methods during the measurement phase or they can be acquired from the equipment data sheets that the manufacturer provides \cite{dobkin2011rf}. }

 In the following, we extend the narrowband model of \cite{mi2017massive} to FBMC. Random calibration errors are considered to be independent at different subcarriers with a constant gain over each subcarrier band. Consequently, these errors can be transformed into the time domain and modeled by a calibration error impulse response at the BS transmit RF chain at any given antenna $i$, i.e., $c_{{\rm t},i}[l]={\color{black}\BDFT}^{-1}\{[\xi_{{\rm t},i}^0e^{j \phi_{{\rm t},i}^{0}}, ..., \xi_{{\rm t},i}^{M-1} e^{j \phi_{{\rm t},i}^{M-1}}] \}$,
 similarly, $c_{{\rm r},i}[l]={\color{black}\BDFT}^{-1}\{[\xi_{{\rm r},i}^0e^{j \phi_{{\rm r},i}^{0}}, ..., \xi_{{\rm r},i}^{M-1} e^{j \phi_{{\rm r},i}^{M-1}}] \}$.
 Accordingly, uplink and downlink channels can be obtained as 
\be
h^{\rm u}_{\color{black} k,i}[l] =  h_{\textcolor{black}{\color{black} k,i}}[l] \star c_{{\rm r},i}[l],
\ee
and
\be
h^{\rm d}_{\textcolor{black}{\color{black} k,i}}[l] = c_{{\rm t},i}\textcolor{black}{[l]} \star h_{\textcolor{black}{\color{black} k,i}}[l],
\ee
respectively. Here, $h_{\textcolor{black}{\color{black} k,i}}[l]$ represents the propagation channel with the PDP defined in Section \ref{sec:flatfbmc}. \textcolor{black}{It is worth noting that the value of errors varies with antenna and subcarriers, but in the case of using equipment with the same configuration and feature, these errors can be considered to follow the same distribution.
}

\subsection{Cell-based Case}
Here, we present an analysis in the presence of imperfect CSI and reciprocity calibration. The precoder with imperfect CSI is obtained by substituting $H^{\rm d}_m({\textcolor{black}{\color{black} k,i}})$ with $\hat{H}^{\rm u}_m({\color{black} k,i}) = H^{\rm u}_m({\color{black} k,i}) + \Delta H^{\rm u}_m({\color{black} k,i})$ in \eqref{eq:combiners}. Therefore, elements of $\bD_m$ in MRT, become
\be
\hat{D}_m^{k,k} = \sum_{i=0}^{N-1}|H_m^{\rm u}({\color{black} k,i})+\Delta H_m^{\rm u}({\color{black} k,i})|^2.
\ee
 Assuming  uncorrelated estimation errors and channel gains, by the law of large numbers, in the asymptotic regime, $\hat{D}_m^{k,k}$ converges to
\begin{align}\label{h_hat_expansion}
 &N \big(\mathds{E}\{|H_m^{\rm u}({\color{black} k,i})|^2\}+\mathds{E}\{|\Delta H_m^{\rm u}({\color{black} k,i})|^2 \big) \nonumber \\ &~~~= N\big(\mathds{E}\{|H_m({\textcolor{black}{\color{black} k,i}})|^2\}+\mathds{E}\{\xi_{{\rm t},i}^m e^{j\phi_{{\rm t},i}^m} \xi_{{\rm t},i}^m e^{-j\phi_{{\rm t},i}^m}\}+\sigma_{\rm ef}^2\big) \nonumber\\
 &~~~=N \textcolor{black}{\beta_k}+N\sigma^2_{\rm ef},
\end{align}
\textcolor{black}{where the variance of $\xi_{{\rm t},i}^m$ is considered to be negligible relative to the variance of channel taps and, hence, is ignored according to models in \cite{mi2017massive}. } \textcolor{black}{Similarly, it can be shown that $(\hat{\bH}_m^{\rm u})^{\rm H} \hat{\bH}_m^{\rm u}$ converges to $\hat\bD_m$. } Therefore, in the asymptotic regime the MRT \textcolor{black}{and ZF precoders have similar performance}.
\color{black} Hence, from \eqref{eq:equivalent-channel}, the equivalent channel between the UE $k$ and the precoder input intended for the $k'$th user over the subcarrier band $m$ converges to 
\begin{align}
        h_{k,k',m}^{\rm (eqvlt)}[l] =&\frac{\textcolor{black}{\sqrt{q_{k'}}}}{N(\textcolor{black}{\beta_{k'}}+{\sigma}_{\rm ef}^2)}\sum_{i=0}^{N-1}  \big(\hat{H}_m({{\color{black} k',i}})\big)^* \xi_{{\rm r},i}^m e^{j \phi_{{\rm r},i}^m}  \nonumber \\ 
        &~~~~~~~~~~~~~~~~~~~~~~~\times\big(h_{k,i}[l]\star c_{{\rm t},i}[l]\big). \label{eq:equivalent_channel_IP}
\end{align}
 Moreover, for large values of $N$, \eqref{eq:equivalent_channel_IP}, reduces to
\begin{equation}
h_{k,k',m}^{\rm (eqvlt)}[l] = \frac{\textcolor{black}{\sqrt{q_{k'}}}}{\textcolor{black}{\beta_{k'}}+{\sigma}_{\rm ef}^2}    \mathds{E} \big\{ \xi_{{\rm r},i}^m e^{j \phi_{{\rm r},i}^m} \big(\hat{H}_m({\textcolor{black}{\color{black} k',i}})\big)^*  h_{\textcolor{black}{\color{black} k,i}}[l] \star c_{{\rm t},i}[l] \big\}.
\end{equation}
 Assuming independent channels for different users, and uncorrelated channel estimation errors, one will find that
 \begin{align}
&h_{k,k',m}^{\rm (eqvlt)}[l] = \frac{\textcolor{black}{\sqrt{q_{k'}}}}{\textcolor{black}{\beta_{k'}}+{\sigma}_{\rm ef}^2}    \mathds{E} \big\{ \xi_{{\rm r},i}^m\big\} \mathds{E} \big\{ e^{j \phi_{{\rm r},i}^m}\big\} \nonumber\\&~~~~~~~~~~~~~~~~
\times \mathds{E} \big\{\big(\hat{H}_m({\textcolor{black}{\color{black} k',i}})\big)^*  h_{\textcolor{black}{\color{black} k,i}}[l] \big\}\star \mathds{E} \big\{ c_{{\rm t},i}[l] \big\}.
\end{align}
 Setting $\lambda =  \mathds{E} \{\xi_{{\rm r},i}^m \} \mathds{E} \{ e^{j \phi_{{\rm r},i}^m}\}$, we note that $\mathds{E} \{c_{{\rm t},i}[l]\}$ is the time domain representation of the calibration error which converges to an impulse with magnitude $\lambda$. {\color{black} While calibration error is not the same for different subcarriers and antennas, it tends to a scaling factor that is the same for all the subcarriers of each user. This is due to the channel hardening effect of massive MIMO. This scaling factor which is derived from the error statistics, can be simply extracted from the device datasheet, as it is noted in \cite{dobkin2011rf}. Thus, similar to works in \cite{chopra2020blind,Shahabi2019novel,raeesi2018performance,mi2017massive}, in this paper, we consider the same statistics for errors at different antennas.} Additionally, following \cite{hosseiny2021fbmc}, one finds that
\be \label{eq:expectationpdp}
\mathds{E}\big\{ \big(\hat{H}_m({\textcolor{black}{\color{black} k',i}})\big)^* h_{\textcolor{black}{\color{black} k,i}}[l] \big\} = p_{k}[l] e^{j2\pi lm/M} \delta_{kk'}.
\ee
\color{black}

Making use of the above results, one will find that the equivalent channel converges to
\be
h_{k,k',m}^{{\rm (eqvlt)}}[l]  =  \tilde{p}_{m,k}[l]\delta_{kk'},
\label{eq:eq_channel_imperfect_co}
\ee
where 
\be 
\begin{aligned}\label{eq:ColocatedPDP_imperfect}
&\tilde{p}_{m,k}[l]= \Big(\frac{\textcolor{black}{\sqrt{q_{k}}} \lambda^2}{\textcolor{black}{\beta_{k}}+{\sigma}_{\rm ef}^2}\Big) \bar{p}_{m,k}[l].
\end{aligned}
\ee 
This shows that the effects of channel estimation and reciprocity calibration errors converge to their statistics. While the errors are subcarrier dependent, in the asymptotic regime, they average out, converge to the same value, and become frequency independent.  As a result, by modifying our proposed two-stage precoder, it is possible to compensate for the imperfect CSI and calibration error effects.  

From the above results, one may realize that to compensate the imperfection of CSI and calibration error effects, the scaling factor  $\frac{\textcolor{black}{\sqrt{q_{k}}} \lambda^2}{\textcolor{black}{\beta_{k}}+{\sigma}_{\rm ef}^2}$ should be added to the PDP $\bar{p}_{m,k}[l]$. This is equivalent to adding the {\em correction factor} $\frac{\textcolor{black}{\sqrt{q_{k}}} \lambda^2}{\textcolor{black}{\beta_{k}}+{\sigma}_{\rm ef}^2}$ to the designed prefilter. \textcolor{black}{The reciprocity calibration error statistics, following the procedures explained in \cite{dobkin2011rf} can be acquired from the data sheet of the equipment. 
Also, in the absence of accurate calibration reciprocity information at the BS, this scaling factor can be easily estimated during the downlink training phase and compensated at the UE end. We note that downlink training is required for synchronization purposes in massive-MIMO systems \cite{jacobsson2019timing} and, thus, these pilots may be used to estimate the correction gain factor.}
It is worth noting that in a case where perfect knowledge of the underlying channel PDP is not available at the base station, it can be estimated from the estimates of the channel as \cite{hosseiny2021fbmc}
\be \label{eq:pdp_approx}
\hat{p}_k[l]=\frac{1}{N} \sum_{i=0}^{N-1}|\hat{h}_{\color{black} k,i}[l]|^2.
\ee
A comparison of different compensation approaches is provided in Section \ref{sec:simulation}, through numerical results

\color{black}
\color{black}

\subsection{Cell-free/Distributed Antenna Case}
\label{sec:calibration_cell_free}
Substituting the imperfect channel estimates in \eqref{eq:equivalent-channel-cellfree}, we obtain an estimate of the equivalent channel between UE $k$ and precoder input of user $k'$ at subcarrier $m$ as \be\label{eq:cell_ResCh_MMSE}
 \hat{h}_{k,k',m}^{\rm (eqvlt)}[l]  = \sum_{i=0}^{N-1} \sqrt{q_{\color{black} k',i}}(\hat{P}_m^{\color{black} k',i})^* \hat{h}^{\rm u}_{\color{black} k,i}[l].
\ee

Following the same line of derivations as in \cite{ngo2013energy}, as the number of antennas, $N$, grows large, the precoders converge to $\bP_m={\rm diag}([\sum_{i=0}^{N-1}\beta_{0,i}, ..., \sum_{i=0}^{N-1}\beta_{K-1,i}])^{-1} \bH_m$.
Hence,
\begin{align}\label{eq:cell_h_eqvlt}
 &\hat{h}_{k,k',m}^{\rm (eqvlt)}[l] \! \rightarrow \! \frac{1}{\sum_{i=0}^{N-1}\beta_{\color{black} k',i}+N{\sigma}_{\rm ef}^2\color{black}} \sum_{i=0}^{N-1}  \sqrt{q_{\color{black} k',i}}\;\mathds{E} \big\{ \xi_{{\rm r},i}^m e^{j \phi_{{\rm r},i}^m} \nonumber \\
 &~~~~~~~~~~~~~~~~~~~~\times \big(\hat{H}_m({\textcolor{black}{\color{black} k',i}})\big)^*  \hat{h}^{\rm u}_{\textcolor{black}{\color{black} k,i}}[l]\big\}  .
 \end{align} 
Also, assuming independent channel responses and uncorrelated estimation errors, one will find that 
\begin{align} \label{eq:MMSE_EqCh_cell}
&\mathds{E} \big\{ (\hat{H}_m^{\color{black} k',i})^*  \hat{h}^{\rm u}_{\color{black} k,i}[l] \big\}= \sum_{l'=0}^{L-1} \mathds{E}\big\{{h}^*_{\color{black} k',i}[l'] {h}^{\rm u}_{\color{black} k,i}[l] \big\} e^{j2\pi l'm/M} \nonumber\\ &~~~~~~~+\sum_{l'=0}^{L-1} \mathds{E}\big\{\Delta h^*_{\color{black} k',i}[l'] \Delta h_{\color{black} k,i}[l] \big\} e^{j2\pi l'm/M} \nonumber\\
&~~~~~~~= \lambda p_{\color{black} k,i}[l]e^{j2\pi lm/M} \delta_{kk'}+{\sigma}_{\rm et}^2 e^{j2\pi lm/M} \delta_{kk'}.
\end{align} 
\color{black}
Substituting \eqref{eq:MMSE_EqCh_cell} in \eqref{eq:cell_h_eqvlt}, leads to 
\begin{align}\label{eq:MMSEinput_est_cell}
& h_{k,k',m}^{\rm (eqvlt)}[l]  \approx \hat{h}_{k,k',m}^{\rm (eqvlt)}[l]-\! \frac{ N\lambda{\sigma}_{\rm et}^2\sum_{i=0}^{N-1}\sqrt{q_{\color{black} k,i}} }{\sum_{i=0}^{N-1} \beta_{\color{black} k,i}+N{\sigma}_{\rm ef}^2}e^{j2\pi lm/M} \delta_{kk'}.
 \end{align}
We propose using this approximation of the equivalent channel to design FSP for the case of cell-free architecture. 
\color{black}As mentioned in the co-located part, the datasheets of equipment can be used to find the error statistics in the correction term following methods in \cite{dobkin2011rf}.

It should be noted that unlike the co-located case, where the reciprocity correction involves the use of a simple scaling factor (the coefficient $(\textcolor{black}{\sqrt{q_{k}}} \lambda^2)/(\textcolor{black}{\beta_{k}}+\sigma_{\rm ef}^2)$), here, the reciprocity correction involves a more complicated correction according to \eqref{eq:MMSEinput_est_cell}. So, the use of pilot symbols that was proposed for reciprocity correction in the co-located case is not applicable to the cell-free case.

  \color{black} \vspace{-1mm} \section{Simulation Results}
 \label{sec:simulation}

In this section, we evaluate our mathematical developments in previous sections by computer simulations.  An FBMC system with $M=64$ subcarriers using a PHYDYAS prototype filter, \cite{bellanger2010fbmc}, with overlapping factor $\kappa=4$ is employed to transmit OQAM (offset quadrature amplitude modulation) symbols. Since all the subcarriers in OFDM experience perfectly flat channels, we set OFDM as a benchmark in our analysis. The 5G channel model, tap delay line-C (TDL-C), \cite{etsi2017138}, is used to obtain the presented results. This model provides a PDP based on a normalized root mean square (RMS) delay spread. The normalized RMS delay spread is randomly scaled for different users in each simulation instance in the range $90$ to $110~{\rm ns}$ for channels with moderate lengths,  \cite{etsi2017138}. This is to address different PDPs between the users and the BS antennas to model practical scenarios.  Additionally, a normalized PDP is assumed, i.e., $\sum_{l=0}^{L-1}p_k[l]=1$ for $k=0,\ldots, K-1$. A sampling frequency of $15.36$~MHz is considered. This results in the subcarrier spacing of $240$~kHz and is inline with 5G NR specifications, \cite{etsi2017138}. To model the magnitude of reciprocity calibration error, we consider $\xi_{{\rm t},i}^m$ and $\xi_{{\rm r},i}^m$ as random variables both with uniform distribution between $0.98$ and $1.02$. Moreover, $\phi_{{\rm t},i}^m$ and $\phi_{{\rm r},i}^m$ are considered as uniformly distributed random variables in the range $[-\frac{2\pi}{9},\frac{2\pi}{9}]$ following measurements in \cite{3GPPmeasurement}. We have obtained our results for 1000 independent realizations of the channel with $K=32$ users. \color{black} A summary of these settings is presented in Table~\ref{tab:my_label}.

\color{black}
\color{black}
\begin{table}[] \color{black}
    \caption{Simulation Parameters}
    \label{tab:my_label}
\begin{center}
\begin{tabular}{ |c|c|c| } 
 \hline
 Number of subcarriers ($M$) & 64 \\ 
\hline
 Prototype Filter & PHYDYAS \\ 
\hline
Channel Model & Tap delay line-C (TDL-C) \\
\hline
Overlapping factor ($\kappa$) & 4 \\
\hline
Number of users ($K$) & 32 \\
\hline
Bandwidth & 15.36~MHz \\
\hline
Pathloss Model  & Cost-Hata \cite{damosso1999cost} \\
\hline
Shadowing Standard Deviation & 8 dB \\
\hline
\end{tabular}
\end{center}
\end{table}

\color{black}
 \subsection{Massive MIMO with Co-located Antennas}
\begin{figure}[t]
		\centering
		\vspace{-2mm}
		\includegraphics[scale=0.65,trim={0 0 0 0},clip]{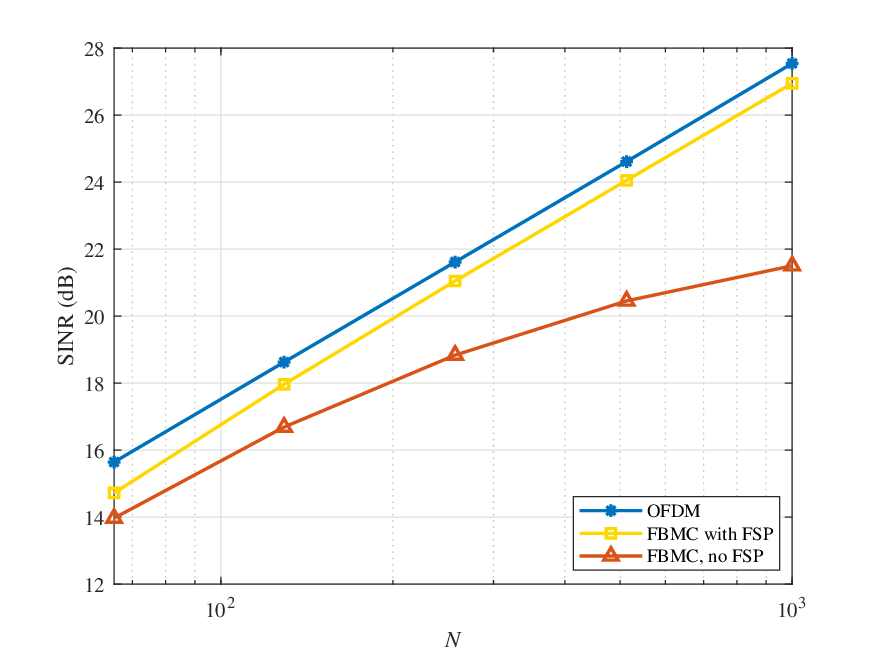}
		\vspace{-8mm}
		\caption{SINR vs. the number of BS antennas, $N$. The FSP design is based on the PDP of the underlying channels. $L_{\rm FSP}=5$.}
		\vspace{-4mm}
		\label{fig:SINRperfect}
	\end{figure}

 \textcolor{black}{This part provides the simulation results on co-located setup.}   \textcolor{black}{We consider an operating SNR of 0 dB.} \textcolor{black}{It is worth noting that employing a power allocation technique leads to canceling out the two coefficients in \eqref{eq:equivalent-channel} and \eqref{eq:pmk[l]}, equivalent to a normalized channel model.}
 To benchmark the results of FBMC against those of OFDM, we present the averaged SINR results in downlink among all the UEs when perfect knowledge of channel is available at the BS. These SINR results are presented  in Fig.~\ref{fig:SINRperfect} as a function of the number of BS antennas, $N$. \textcolor{black}{A ZF precoder is employed at the second stage of  Fig.~\ref{blockdiagram_DL}.}  The FSP design follows the formulation presented in \cite{hosseiny2021fbmc}, for an FSP length $L_{\rm FSP}=5$. \textcolor{black}{While larger $L_{\rm FSP}$ lengths than $5$ lead to an improved performance, $L_{\rm FSP}=5$ balances the complexity, latency and performance.} As seen, SINR results for FBMC follow those of OFDM, with a loss of about \textcolor{black}{$0.7$}~dB, resulting from the limited length of FSP. This gap reduces if $L_{\rm FSP}$ is increased. The results also show a high impact of FSP in improving SINR. For larger values of $N$, the performance gain resulting from \textcolor{black}{use of} FSP can be several decibels; even greater than $5$~dB as $N$ approaches its upper range in  Fig.~\ref{fig:SINRperfect}.
 
 
\begin{figure}[t]
		\centering
		\vspace{-2mm}
  \hspace{-8mm}
		\includegraphics[scale=0.65,trim={0 0 0 0},clip]{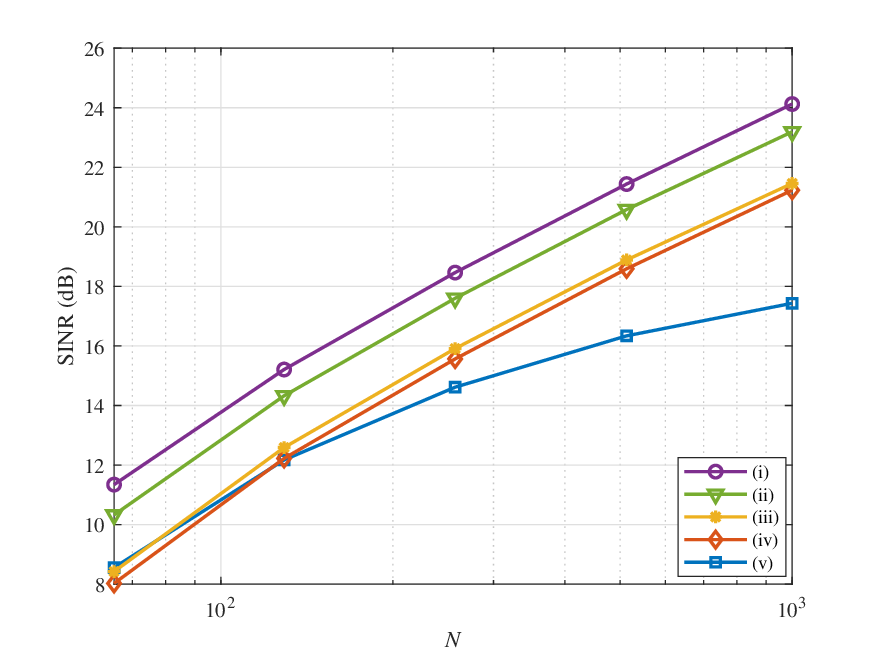}
		\caption{Output SINR vs. the number of BS antennas, $N$. Variety of cases (discussed in the text) are presented to highlight the impact of various effects. $L_{\rm FSP} = 5$. }
		\label{fig:SINRimperfect}
	\end{figure}

In Fig.~\ref{fig:SINRimperfect}, we analyze  the impact of channel reciprocity effects and the performance gain obtained using our  proposed calibration and channel estimation error compensation techniques, presented in Section~\ref{sec:calibration}. The channel estimation method of \cite{Hoss2006:Spectrally} is deployed in the uplink to obtain different users' channel responses.  To obtain reasonably accurate channel estimates, the power allocated to pilot symbols was boosted to a level $10$~dB above the allocated power for data transmission. As it was mentioned before, without reciprocity     calibration at the BS, the scaling factor $\frac{\textcolor{black}{\sqrt{q_{k}}} \lambda^2}{\textcolor{black}{\beta_{k}}+{\sigma}_{\rm ef}^2}$ needs to be estimated at the UE side. Hence, a set of pilots, similar to the ones in \cite{Hoss2006:Spectrally}, are transmitted in the first FBMC block of each downlink packet to obtain the scaling factor for calibration and channel estimation error compensation at the UE side.  As it is shown in Fig.~\ref{fig:SINRimperfect}, pilot aided correction of the scaling factor $\frac{\textcolor{black}{\sqrt{q_{k}}} \lambda^2}{\textcolor{black}{\beta_{k}}+{\sigma}_{\rm ef}^2}$ at the UE leads to a very close performance to that of perfect correction when the calibration information is available at the BS.


	
Comparing the results in Fig.~\ref{fig:SINRperfect} and those in Fig.~\ref{fig:SINRimperfect}, one may observe several dBs loss in performance due to channel estimation error and imperfect channel reciprocity effects. Such losses are in line with those reported for OFDM in the literature, e.g., see \cite{mi2017massive}. To highlight the losses arising from channel reciprocity and channel estimation errors separately, the results that account for each of these effects are separately presented in Fig.~\ref{fig:SINRimperfect}. The roman numbers indicated as legends in Fig.~\ref{fig:SINRimperfect}, and later used in Fig.~\ref{fig:cellfree_imperfectCSI_DL} as well, refer to the following cases.
\begin{enumerate}[label=(\roman*)]
\item FSP is applied, perfect CSI is assumed, in presence of channel reciprocity errors that remain uncompensated.
\item FSP is applied, CSI is estimated, there is no channel reciprocity error.
\item FSP is applied, CSI is estimated, in presence of channel reciprocity errors that are perfectly compensated.
\item FSP is applied, CSI is estimated, in presence of channel reciprocity errors that are compensated using downlink pilot symbols.
\item FSP is not applied, CSI is estimated, in presence of channel reciprocity errors that remain uncompensated.
\end{enumerate}

 \subsection{Cell-free Massive MIMO}
 \label{sec:cellfresimulation}
 In this subsection, we evaluate the performance of our proposed techniques in Sections~\ref{sec:cell_free} and \ref{sec:calibration} through simulations. To this end, we consider a cell-free massive MIMO setup with APs that are located on a regular grid in an area of $2\times2$ square kilometers, each equipped with $4$ antennas. We deploy the wrap-around technique of \cite{bjornson2019making} to imitate an infinite area and thus, avoid boundary effects.
Small-scale fading is simulated based on the TDL-C model, \cite{etsi2017138}. The large-scale fading coefficients, $\beta_{i,k}$, are calculated using the COST Hata model, \cite{damosso1999cost}, i.e., 
\be
10\text{log}_{10}(\beta_{i,k}) = -135-35\text{log}_{10}(d_{i,k})-\mathcal{X}_{i,k},
\label{eq:beta}
\ee 
where $d_{i,k}>10~{\rm m}$ is the distance between a given user $k$ and antenna $i$ in kilometers and $\mathcal{X}_{i,k} \sim \mathcal{CN}(0, \sigma^2_{\mathcal{X}})$  represents shadowing effect with $\sigma^2_{\mathcal{X}}=8$ dB. Variance of noise is obtained as $\sigma^2_{\eta} = \mathcal{K} \times \kappa_{\rm B} \times  B \times {\rm NF}$,
where $\mathcal{K}$, $\kappa_{\rm B}$, $B$, and NF are temperature in kelvin, Boltzmann constant, bandwidth, and noise figure, respectively. We consider $\mathcal{K}=290$~K, $\kappa_{\rm B} = 1.3 \times 10^{-23}$~J/K, $B=20$~MHz, and NF~$=9$ dB. Maximum transmit power of each antenna is set to \textcolor{black}{$250$~mW}.

\begin{figure}
		\centering
  \vspace{-1.8mm}
		\includegraphics[scale=0.65,trim={0 0 0 0},clip]{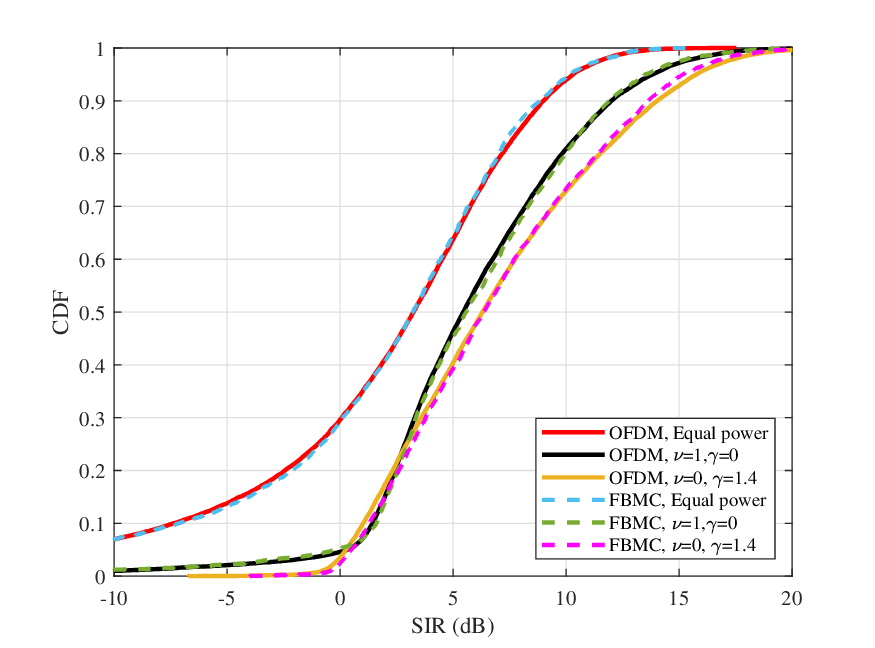}
		\caption{Comparison of the empirical CDF of OFDM and FBMC while using maximum power and power allocation.}
		\label{fig:cdf}
	\end{figure}

\color{black}
Fig.~\ref{fig:cdf} illustrates the cumulative distribution function (CDF) of the signal-to-interference ratio (SIR) performance for FBMC- and OFDM-based cell-free MIMO setup with \textcolor{black}{$100$} APs in the area of $2\times 2$ ${\rm km}^2$.
The results show that power allocation leads to more stable values (i.e., less variation) in SIRs. This is inline with the previous results on OFDM in the literature \cite{nikbakht2020uplink}. Our results, here, confirm that the same is true for FBMC, and power allocation has almost the same impact on both OFDM and  FBMC. Following the recommendations made in the literature,  \cite{simonsson2008uplink,yates1995framework} and \cite{nikbakht2020uplink}, in the rest of this section, we consider the fractional power allocation with $\nu=0.6$ and $\gamma = 1.2$. It is worth noting that, power allocation leads to a higher average SINR by limiting the multiuser effect.

\begin{figure}
		\centering		\includegraphics[scale=0.63,trim={0 0 0 0},clip]{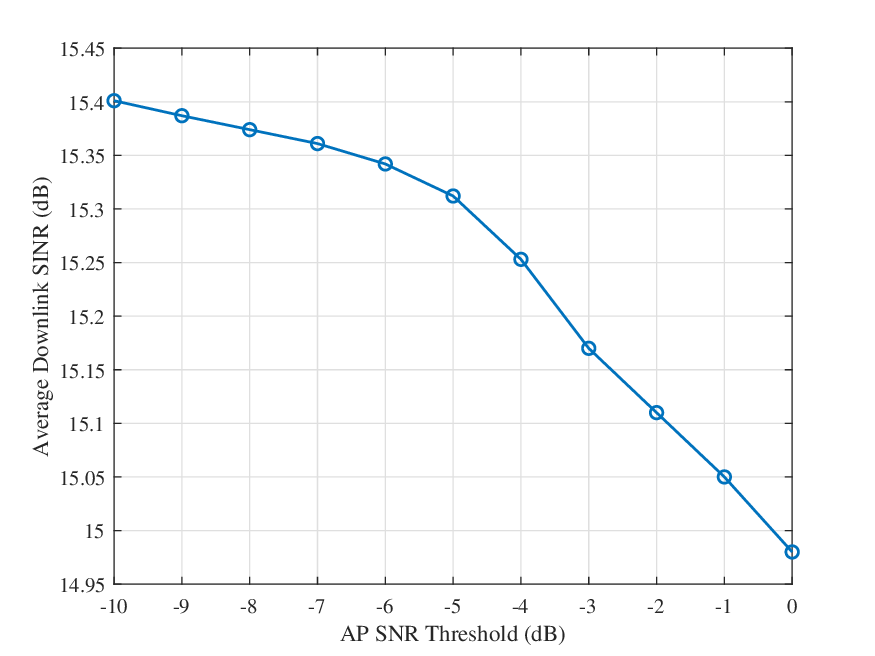}
		\caption{Variation of output SINR vs. AP selection threshold.}
		\label{fig:cellfree_thvsSINR}
\end{figure}

 \color{black}
 In Fig.~\ref{fig:cellfree_thvsSINR}, we analyze the SINR performance of our proposed AP selection technique in Section~\ref{sec:cell_free} as a function of the SNR threshold \textcolor{black}{for $N_{\rm AP}=100$. As explained in Section~\ref{sec:cell_fee_AP_selection}, the proposed antenna selection method selects an AP to be in $\mathcal{B}_k$, if that AP gets a signal with an SNR larger than the threshold. As Fig.~\ref{fig:cellfree_thvsSINR} shows,} increasing the SNR threshold beyond \textcolor{black}{$-5$~dB}, leads to a faster SNR drop, with a noticeable change in the slope of the drop. 
 \textcolor{black}{Similarly, the number of effective antennas starts to drop by increasing the threshold.}
 Thus, a compromise choice of threshold that reduces the number of selected APs while keeping the achievable SINR at an acceptable level should be selected. For the remaining results here, we set the threshold equal to $-5$~dB. Compared to lower threshold values, e.g., $-10$~dB, this incurs a fraction of $1$ dB loss in SINR, while reducing the number of active APs by a factor of one half or smaller.

	\begin{figure}
		\centering
		\includegraphics[width=92mm,scale=0.545,trim={0 0 0 0},clip]{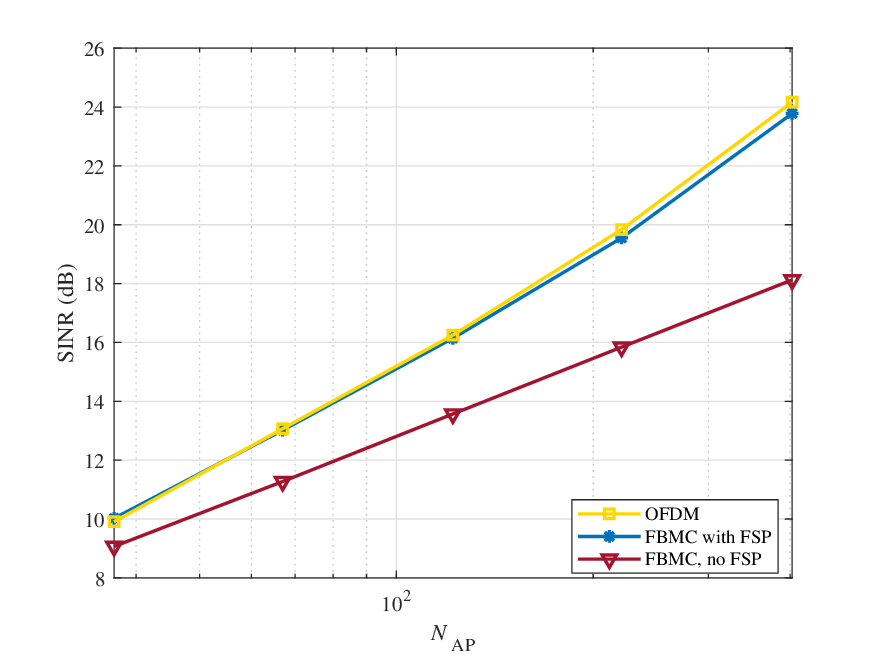}
		\caption{Output SINR vs. number of APs, $N_{\rm AP}$, for FSP design  $L_{\rm FSP}=5$.}
		\label{fig:perfectCSI_cellfree_ZF}
	\end{figure}
With the above considerations, the performance of our proposed precoding technique \textcolor{black}{for cell-free massive MIMO} is evaluated in  Fig.~\ref{fig:perfectCSI_cellfree_ZF}. Similar to massive MIMO with co-located antennas, we set OFDM as a benchmark to show the channel flattening capability of the proposed precoder. The CSI is assumed to be perfectly known and the FSP length is chosen as $L_{\rm FSP}=5$. A ZF precoder is employed at the first stage of  Fig.~\ref{blockdiagram_DL}. Here also it is observed that our proposed method brings some performance improvement when compared to the conventional single-tap precoding. As for the comparison with OFDM, here, FBMC leads to a comparable performance. The performance gap observed in Fig.~\ref{fig:SINRperfect} is not seen here.  We may also note that the SINR ranges seen in Fig.~\ref{fig:SINRperfect} are significantly higher than those in Fig.~\ref{fig:perfectCSI_cellfree_ZF}. This is related to the fact that in the cell-free case, the number of effective antennas that contribute to the antennas processing gain is significantly lower than those in the co-located antennas system.



\begin{figure}
		\centering		\includegraphics[scale=0.63,trim={0 0 0 0},clip]{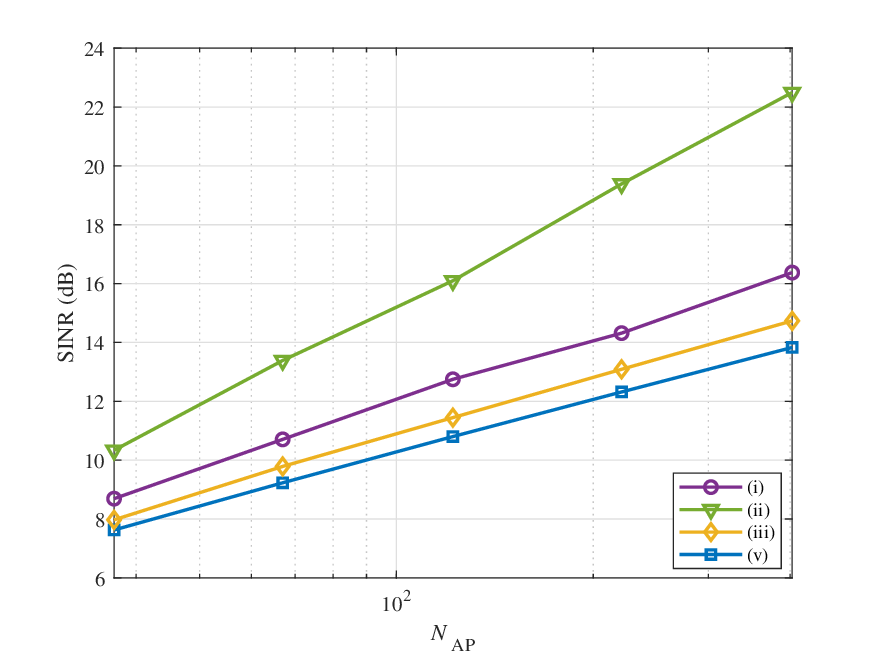}
		\caption{Output SINR vs. number of APs, $N_{\rm AP}$, for FSP design  $L_{\rm FSP}=5$. For the details of legends, refer back to the text at the end of Section VI-A.}
		\label{fig:cellfree_imperfectCSI_DL}
\end{figure}

The above observation is carried over in Fig.~\ref{fig:cellfree_imperfectCSI_DL} as well, where we study the effects of imperfect CSI and reciprocity calibration errors on the performance of the proposed precoder. As one would expect, the results here also show the positive impact of FSP and the reciprocity correction factor. However, in comparison with the results in Fig.~\ref{fig:SINRimperfect}, the use of FSP in cell-free massive MIMO brings a limited performance gain. As discussed in Section~\ref{sec:cell_fee_AP_selection}, and also emphasized above, the  limited number of contributing/effective antennas accounts for the observations made here when the results are compared with those in Fig.~\ref{fig:SINRimperfect}. It is also noted that, here, the presence of channel reciprocity has a significant impact on the SINR results. Moreover, as noted at the end of Section \ref{sec:calibration_cell_free}, the case of reciprocity compensation through downlink pilots is not applicable here.



\color{black}
 \vspace{-1mm}\section{Conclusion}
\label{sec:conclusion}
In this work, we developed a practical precoding method for the downlink of FBMC-based massive MIMO in co-located and distributed antenna setups. Theoretical results that show the impact of channel estimations error and reciprocity mismatch in uplink and downlink radio chains were developed. The proposed method includes a two-stage precoder. The first stage of the precoder applies a fractionally spaced equalizer (FSP) for flattening/equalizing the channel across each subcarrier band. The second stage is a conventional precoder. \textcolor{black}{In this paper, we used a ZF precoder.} The downlink of an FBMC-based cell-free/distributed architecture was formulated, and a precoding method has been proposed. Then, we proposed an access point (AP) selection technique and power allocation method for this distributed scenario. We also studied the theoretical impact of calibration and channel estimation errors and proposed compensation approaches to reduce the effects of imperfections. We showed that in co-located massive MIMO, these errors could trivially be obtained by sending a pilot signal and compensating through a single scaling factor that is similar for all the subcarriers, thanks to the channel hardening effect. In the cell-free setup, the correction term was added to the precoder design. Simulation results that corroborate our theoretical findings were also presented. 

\vspace{-2mm}

\ifCLASSOPTIONcaptionsoff
  \newpage
\fi

\bibliography{ref.bib}
\bibliographystyle{IEEEtran}
%


\end{document}